\documentclass{jfm}
\pdfoutput=1
\usepackage{natbib}
\usepackage{amsmath}
\usepackage{subcaption}
\usepackage{booktabs}
\usepackage{graphicx}
\usepackage[percent]{overpic}
\usepackage{color}
\usepackage{tikz}
\usepackage{flafter}

\usepackage{ulem}

\begin{document}

\def\do#1{{d\over d#1}}
\def\dt#1#2{{d#1\over d#2}}
\def\dpo#1{{\partial\over\partial#1}}
\def\dpt#1#2{{\partial#1\over\partial#2}}
\def\dtpt#1#2{{\partial^2#1\over\partial{#2}^2}}
\def\dtpth#1#2#3{{\partial^2#1\over\partial#2\partial#3}}

\def\vx{\boldsymbol{x}}
\def\vy{\boldsymbol{y}}
\def\vr{\boldsymbol{r}}
\def\vk{\boldsymbol{k}}
\def\vu{\boldsymbol{u}}
\def\va{\boldsymbol{a}}
\def\vb{\boldsymbol{b}}
\def\vc{\boldsymbol{c}}
\def\vI{\boldsymbol{I}}
\def\vA{\boldsymbol{A}}
\def\ven{\hat{\boldsymbol{n}}}
\def\hf{\hat{\boldsymbol{f}}}
\def\hG{\hat{\boldsymbol{G}}}
\def\vpsi{\boldsymbol{ \psi}}
\def\vomega{\boldsymbol{ \omega}}
\def\hvpsi{\hat{\mbox{\boldsymbol{ \psi}}}}
\def\hvo{\hat{{\mbox{\boldsymbol{ \omega}}}}}
\def\hvu{\hat{\vu}}
\def\cS{{\cal S}}
\def\cN{{\cal N}}
\def\cH{{\cal H}}
\def\cE{{\cal E}}
\def\cC{{\cal C}}
\def\cO{{\cal O}}

\def\R{\mathbb{R}}

\newcommand{\intdiff}{\mathop{}\!\mathrm{d}}

\newcommand{\blackline}{\raisebox{2pt}{\tikz{\draw[-,black,solid,line width = 1pt](0,0) -- (5mm,0);}}}
\newcommand{\redline}{\raisebox{2pt}{\tikz{\draw[-,red,solid,line width = 1pt](0,0) -- (5mm,0);}}}
\newcommand{\blueline}{\raisebox{2pt}{\tikz{\draw[-,red,blue,line width = 1pt](0,0) -- (5mm,0);}}}
\newcommand{\reddash}{\raisebox{2pt}{\tikz{\draw[-,red,dashed,line width = 1.0pt](0,0) -- (5mm,0);}}}
\newcommand{\orangedash}{\raisebox{2pt}{\tikz{\draw[-,orange,dashed,line width = 1.0pt](0,0) -- (5mm,0);}}}
\newcommand{\bluedash}{\raisebox{2pt}{\tikz{\draw[-,blue,dashed,line width = 1.0pt](0,0) -- (5mm,0);}}}
\newcommand{\greendot}{\raisebox{2pt}{\tikz{\draw[-,green!40!gray,dotted,line width = 1.0pt](0,2) -- (5mm,2);}}}
\newcommand{\reddot}{\raisebox{2pt}{\tikz{\draw[-,red,dotted,line width = 1.0pt](0,2) -- (5mm,2);}}}
\newcommand{\bluedotdash}{\raisebox{2pt}{\tikz{\draw[-,blue,dashdotted,line width = 1.0pt](0,2) -- (5mm,2);}}}
\newcommand{\greendotdash}{\raisebox{2pt}{\tikz{\draw[-,green!40!gray,dashdotted,line width = 1.0pt](0,2) -- (5mm,2);}}}
\newcommand{\greendashdotdot}{\raisebox{2pt}{\tikz{\draw[-,green!40!gray,dashdotdotted,line width = 1.0pt](0,2) -- (5mm,2);}}}
\newcommand{\browndashdotdot}{\raisebox{2pt}{\tikz{\draw[-,brown!40!gray,dashdotdotted,line width = 1.0pt](0,2) -- (5mm,2);}}}
\newcommand{\greendensedashdotdot}{\raisebox{2pt}{\tikz{\draw[-,green!40!gray,densely dashdotdotted,line width = 1.0pt](0,2) -- (5mm,2);}}}
\newcommand{\cyandash}{\raisebox{2pt}{\tikz{\draw[-,cyan,dashed,line width = 1.0pt](0,0) -- (5mm,0);}}}

\shorttitle{Spherical region of turbulence} 
\shortauthor{K. Y. et al} 

\title{Dynamics and decay of a spherical region of turbulence in free space}

\author
 {
 Ke Yu\aff{1}
  \corresp{\email{kyu2@caltech.edu}},
  Tim Colonius\aff{2},
  D.I. Pullin\aff{1}
  \and
  Gr{\'e}goire Winckelmans\aff{3}
  }

\affiliation
{
\aff{1}
Graduate Aeronautical Laboratories, California Institute of Technology,
Pasadena, CA 91125, USA
\aff{2}
Department of Mechanical and Civil Engineering, California Institute of Technology,
Pasadena, CA 91125, USA
\aff{3}
Institute of Mechanics, Materials and Civil Engineering, Universit{\'e} catholique de Louvain, Louvain-la-Neuve, Belgium
}

\maketitle
\begin{abstract}
{We perform direct numerical simulation (DNS) and large eddy simulation (LES) of an initially spherical region of turbulence evolving in free space.  The computations are performed with a lattice Green's function method, which allows the exact free-space boundary conditions to be imposed on a compact vortical region. LES simulations are conducted with the stretched vortex sub-grid stress model.  The initial condition is spherically windowed, isotropic homogeneous incompressible turbulence.   We study the spectrum and statistics of the decaying turbulence and compare the results with decaying isotropic turbulence, including cases representing different low wavenumber behavior of the energy spectrum (i.e. $k^2$ versus $k^4$).
At late times the turbulent sphere expands  with both mean radius and  integral scale showing similar time-wise growth exponents.
The low wavenumber behavior has little effect on the inertial scales, and we find that decay rates follow \citet{saffman1967large} predictions in both cases, at least until about $400$ initial eddy turnover times.  The boundary of the spherical region develops intermittency and features ejections of vortex rings.  These are shown to occur at the integral scale of the initial turbulence field and are hypothesized to occur due to a local imbalance of impulse on this scale.
}
\end{abstract}

\section{Introduction}
We study the evolution of a compact region of turbulence in an otherwise unbounded domain.  Our interest in this novel flow started in seeking a simple validation case for a large-eddy simulation (LES) model in free space.  In past work, validations have involved canonical flows in wall-bounded or periodic domains.  In free space, and without artificial forcing, there are no mechanisms to sustain turbulence, and it will decay in time.  Of possible initial flow fields, two of relevance here are a collection of one or more vortex rings and a random initial condition.

Initial conditions comprising a collection of vortex rings are readily created in both experiments and simulations. Recent experiments have studied  the generation of  turbulence  through vortex-ring collisions
\citep{matsuzawa2019realization} and have evaluated hypothesized mechanisms of turbulence self-sustenance \citep{mckeown2018cascade}.
The resulting turbulence occurs through a complex process of instabilities, vortex interaction, and reconnection \citep{lim1992instability}. Simulations are computationally intensive since distinct turbulent and laminar regions will occur, and the associated vortex-ring Reynolds numbers must be sufficiently high for transition to occur. In contrast, a random initial condition is computationally
simple and turbulent Reynolds numbers can easily be reached (even in DNS), but a disadvantage is that there will be an initial transient period that, while governed by the Navier-Stokes equations, will not be associated with physical turbulence.  As a compromise, we manufacture an initial condition by first generating isotropic homogeneous turbulence (IHT) in a periodic domain, and initializing a free-space {\it cloud} of turbulence by tiling the periodic solution in space and windowing it with an indicator function that falls to zero outside a sphere of radius $R$, which can be varied compared to the initial scales in the IHT.

Several features of the evolution of such a spherical region of turbulence in free space are of theoretical interest.  In IHT, the evolution of the largest scales is governed by the initial conditions, or, in the case of forced IHT, by the forcing scheme. For example, spectra with low wavenumber that asymptotes as $k^2$ \citep{saffman1967large} and $k^4$ \citep{batchelor1956large} can be contrived \citep{chasnov1995decay,ishida2006decay,davidson2010decay}. Indeed, the same is true for the spherical turbulence cloud, where, unlike IHT, the largest scales can subsequently grow and, as we will show, different behavior is obtained.
The spherical region of turbulence also exemplifies the localized turbulence region introduced in \citet{phillips1956final}, where the final viscous stage of the evolution was studied theoretically.

A second motivation concerns the emergence of coherent structures.  IHT is devoid of large-scale instabilities (typically associated with shear, buoyancy, or other imposed forces) that give rise to important classes of coherent structures.  However, as we show, the same is not the case for the spherical cloud --- we observe the formation of coherent vortex rings being ejected near the cloud edge.

A related issue is the interaction of the turbulent flow with the outer irrotational fluid at the turbulent/nonturbulent interface (TNTI).  In recent work, TNTIs have been experimentally and numerically studied in shear layers and in numerically-constructed shear-free interfaces \citep{ wolf2013,de2013multiscale,da2014characteristics}.  The spherical cloud of turbulence also exhibits a TNTI, and may prove a useful source of data for further study, though in the present paper we do not investigate its behavior in detail.

In this paper, we present DNS and LES simulations for the spherical region of turbulence and examine the resulting energy spectrum and its decay, and the ejection of vortex rings from its periphery.
In \S~\ref{sec:numerical_method} we present the numerical method used to solve the incompressible Navier-Stokes equations in free space, and introduce the turbulence model used in the LES.
In \S~\ref{sec:initial_conditions}, the initial conditions are discussed in detail, including recipes for generating $k^2$ and $k^4$ spectra.
In \S~\ref{sec:viz}, DNS and LES results are used to visualize the evolution of the turbulence field.
In \S~\ref{sec:statistics}, we use statistical measures to characterize the decay. We also show LES calculations agree well with the DNS in these measures. In \S~\ref{sec:longterm_statistics}, LES is used to study the long-term evolution of the turbulence cloud.
Finally, in \S~\ref{sec:vortex_rings}, we discuss one distinctive feature in the long-term evolution, the ejection of vortex rings, and we conjecture about the relationship between the initial condition and the scale of the ejections.  A brief summary of the main conclusions is given in \S~\ref{sec:conclusions}.
\section{Numerical method}\label{sec:numerical_method}

\subsection{The fast lattice Green's function method}\label{sec:lgf}

We solve the incompressible (constant density and viscosity) Navier-Stokes equations in an unbounded three-dimensional space.  A second-order mimetic finite volume scheme based on the fast lattice Green's function (LGF) method recently developed by \citet{liska2016fast}
is applied. The method enforces the divergence-free constraint by solving the associated (discrete) Poisson equation on a formally infinite grid using the LGF technique.

For flows with a compact vorticity field, the infinite lattice can be truncated to a finite region that adapts to the local flow according to a threshold value on the vorticity where sources to the Poisson equation are finite.  The solution can be reconstructed at any position, but is only done at those lattice points needed for the next time step.
This method is ideal for incompressible external flow simulations for the following two reasons: first, it yields more accurate solutions since the exact far-field boundary condition is embedded in LGFs and no artificial outflow boundary condition is imposed; second, it is efficient because the computation domain is snug around the vortical region.

Other building blocks of the scheme include: a third-order Runge-Kutta scheme for the time integration; an analytical integrating factor technique for the viscous term which has the advantage of neither introducing discretization errors nor imposing stability constraints on the time step.  The scheme is parallelized and has been extensively validated by comparison with exact solutions and grid refinement studies \citep{liska2014parallel}.  The DNS simulation reported here uses a maximum of around $2\times 10^9$ computational cells running on $1,500$ cores.

\begin{figure}
    \centering
    \begin{overpic}[width=.95\textwidth]{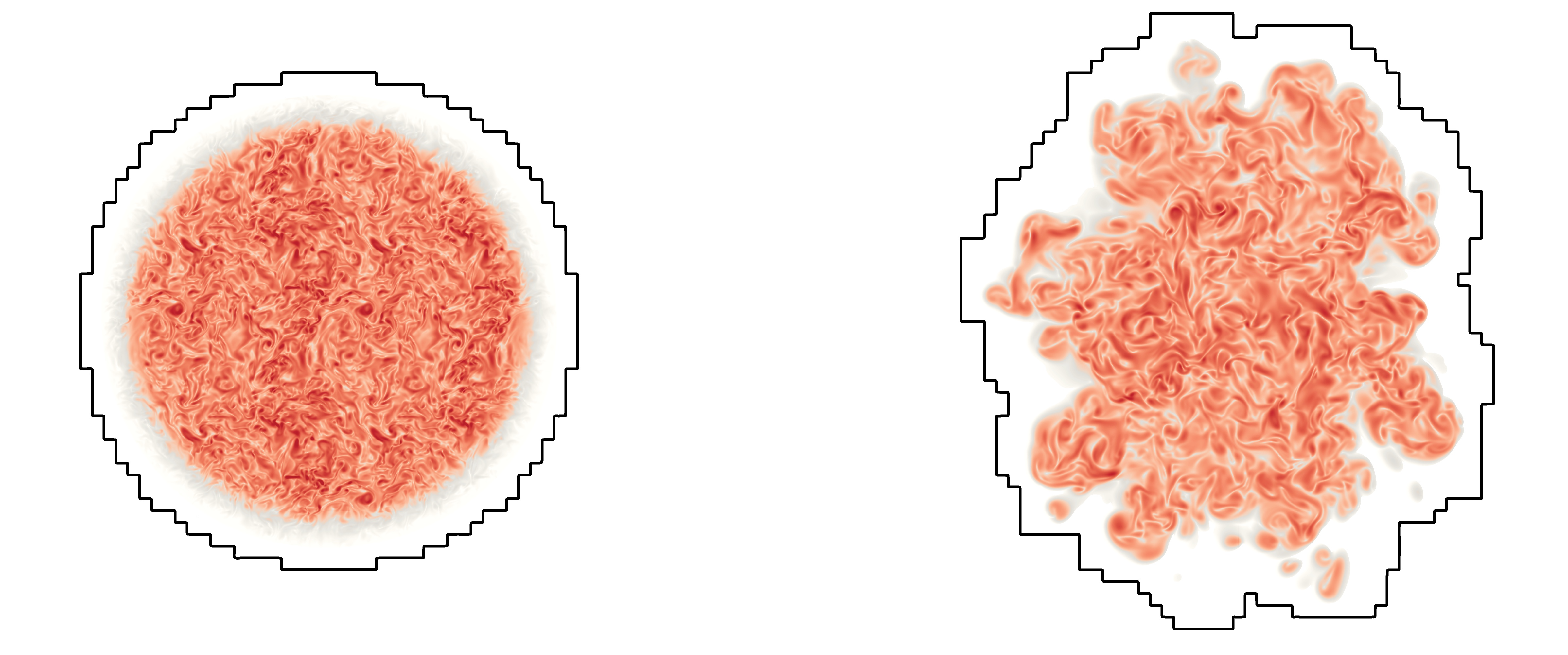}
        \put(0,37) {\small $t/t_\ell=0$}
        \put (53,37) {\small $t/t_\ell=8.6$}
    \end{overpic}
    \caption{\small
     Vorticity magnitude in a cross-section through the center at $t/t_\ell=0$ and $t/t_\ell=8.6$, corresponding to the case DNS\_0 from table \ref{tab:simulation_summary}. Black contour lines indicate the spatially adaptive computational domain. The smallest adaptivity unit is a block of $16^3$ computational cells. }\label{fig:IC}

\end{figure}

Figure \ref{fig:IC} shows the setup for the problem at hand. Initially only a spherical region in free space is filled with turbulence (fluid with non-zero vorticity). The spherical region then starts to deform, evolve and decay.  This method utilizes fixed local cell size but is able to spatially adapt with the vortical areas by adding or removing blocks of computation cells (black contour lines in figure \ref{fig:IC}).  All simulations here are conducted with a spatial adaptive threshold $\epsilon_{\text{supp}}$ equal to $10^{-5}$ defined in \citet{liska2016fast}. Because of the spatial adaptivity, the total number of computation cells varies through one simulation.

\subsection{Stretched vortex sub-grid stress (SGS) model }\label{sec:stretched_vortex}

The LES solutions we report rely on the  stretched vortex SGS model (SVM) \citep{chung2009large}.  The LES solves the filter-averaged Navier-Stokes equations
\begin{align}\label{eq:N-S-LES}
    \frac{\partial \widetilde{\vu}}{\partial t}+
    \widetilde{\widetilde{\vu}   \cdot \nabla \widetilde{\vu}}
    =- \nabla \widetilde{P}+\nu \nabla^{2} \boldsymbol{\widetilde{\vu}} - \nabla \cdot \widetilde{\mathsfbi{T}},
    \\
    \nabla \cdot \boldsymbol{\widetilde{\vu}}=0,
\end{align}
where $\widetilde{\mathsfbi{T}} = \widetilde{\vu\otimes \vu} - \widetilde{\widetilde{\vu} \otimes \widetilde{\vu}}$ is the
SGS tensor.
The model assumes
that the subgrid motions for a single computational cell are dominated by small vortices in
a direction $\boldsymbol{e^v}$ that is aligned with the principle eigenvector of the
resolved strain rate tensor. Then the SGS tensor is given by
\begin{align}
    \widetilde{T}_{i j}=K \left(\delta_{i j}-e_{i}^{v} e_{j}^{v}\right), \\
    K=\int_{k_{c}}^{\infty} E(k) \intdiff k=
    \mathcal{K}_{0}' \Gamma\left[-1 / 3, \kappa_{c}^{2}\right] / 2, \label{eg:sgs-kinetic}
\end{align}
where $K$ is the subgrid kinetic energy, $k_c = \pi / \Delta_x= \pi / \Delta_y= \pi / \Delta_z$ with $\Delta_{(\cdot)}$ being the cell length, and $\Gamma$ is the incomplete gamma function.  The second equality in equation (\ref{eg:sgs-kinetic}) assumes that SGS vortices are of the stretched-spiral type with spectrum \citep{lundgren1982strained}
\begin{align}
    E(k)=\mathcal{K}_{0} \epsilon^{2 / 3} k^{-5 / 3} \exp \left[-2 k^{2} \nu /(3|\widetilde{a}|)\right] \label{eq:lundgren},
\end{align}
where  $\nu$ is the fluid viscosity
\begin{align}
\widetilde{a}=e_{i}^{v} e_{j}^{v} \widetilde{S}_{i j}, \quad
    \mathcal{K}_{0}'=\mathcal{K}_{0} \epsilon^{2 / 3} \lambda_{v}^{2 / 3}, \quad
    \lambda_{v}=(2 \nu / 3|\widetilde{a}|)^{1 / 2}, \quad \kappa_{c}=k_{c} \lambda_{v},
\end{align}
and $\widetilde{S}_{i j}$ is the resolved strain rate tensor. Finally the constant $\mathcal{K}_{0}'$ in equation (\ref{eg:sgs-kinetic}) is determined by matching the resolved second-order velocity structure function with the prediction from the energy spectrum given by equation~(\ref{eq:lundgren}). Details regarding the efficient evaluation of the aforementioned SGS stress can be found in \citet{voelkl2000physical,chung2009large}. 

The SVM is structure based  and not of the eddy-viscosity type. All model parameters are calculated dynamically using only local information from the resolved-scale field surrounding the grid cell or point where sub-grid stresses are calculated. The SVM keeps track of the actual fluid viscosity and also the subgrid kinetic energy, and will automatically become subdominant to real viscous stresses when the flow is locally resolved.  It has proven robustness and  has been successfully used for studies of decaying turbulence \citep{misra1997vortex},  and wall-resolved LES of channel flow \citep{voelkl2000physical,chung2010direct}, bluff-body flows \citep{cheng2017large,cheng2018large} and Taylor-Couette flow \citep{cheng2020large}.

\subsection{Initial condition}\label{sec:initial_conditions}

The initial condition is generated by spherically windowing a turbulence field from a separate IHT computation with periodic boundary conditions. This field is then tiled in all directions to fill the free space and the velocity field is     multiplied by a smooth window function of the form
\begin{align}\label{eq:window}
    \Phi(r) & = \frac{1}{2}\left[1-\tanh \left( \frac{2 ( r-R)}{\sigma}\right) \right],
\end{align}
where $R$ is the radius of the sphere and $\sigma$ is the width of the transition, whose impact on the results will be assessed.  The forced periodic IHT field is generated using a simple 3D pseudo-spectral code and we  define the domain size to be $B^3$.
A low wavenumber forcing method is applied \citep{huang1994}. The forcing is restricted to modes with wavenumbers $|\boldsymbol{k}|<2.5$ and the magnitude of the forcing is chosen to keep the energy input rate constant, which  would equal to the dissipation rate $\epsilon$ after the forced turbulence becomes stationary. To make sure all IHT flows are fully resolved,  $\epsilon$ is
determined such that $\eta k_{\mathrm{max}} \sim 1.5$, where $k_{\mathrm{max}} = N_s / 2$ is
the maximum wavenumber and $\eta = \left({\nu^{3}}/{\epsilon}\right)^{1 / 4}$ is the Kolmogorov length scale with $\nu$ being the viscosity. We also confirmed the isotropy of the IHT field by verifying that $ \frac{E_{ii}(k)}{E(k)} - \frac{1}{3} \approx 0, \ i=1,2,3$.

Figure \ref{fig:IC} visualizes the initial vorticity field in a cross-section through the center. This corresponds to the case DNS\_0 defined in table \ref{tab:simulation_summary}. Note that black contour lines in figure \ref{fig:IC} are the spatially adaptive computational domain which encompasses the initial voricity field as discussed in \S \ref{sec:lgf}. For LES, the IHT field is spectrally filtered before tiling and windowing. More about initial conditions for LES is discussed in \S~\ref{sec:DNSLES_visualization} when results from LES are presented.

Once an IHT field and resolution are selected (which give an initial turbulence Reynolds number, $\Rey_\lambda$), two non-dimensional parameters characterize the initial condition: $B / R$ and $\sigma / R$.  Table \ref{tab:simulation_summary} summarizes the parameters for all runs studied in this work. The simulation parameters used in the pseudo-spectral code to generate the IHT fields and their statistical characteristics are given in table \ref{tab:IHT}.

\begin{table}
\centering
\begin{tabular}{lrrrrrr}
Name    \qquad& $\Rey_\lambda$ \qquad& $ \sigma / R$ \qquad& $B / R$ \qquad & Spectrum Type\\
\\
DNS\_0 & 122.4              & 0.10  & 1.0       & $2$          \\
LES\_0 & 122.4              & 0.10  & 1.0       & $2$             \\
LES\_IC2   &  122.4         & 0.10  & 1.0       & $4$              \\
LES\_D1    & 122.4          & 0.05  & 1.0       & $2$            \\
LES\_D2    & 122.4          & 0.20  & 1.0       & $2$             \\
LES\_B1    & 122.4          & 0.10  & 0.5       & $2$             \\
LES\_B2    & 122.4          & 0.10  & 2.0       & $2$              \\
LES\_R1   & 76.9            & 0.10  & 1.0       & $2$              \\
LES\_R2   & 45.0            & 0.10  & 1.0       & $2$              \\

\end{tabular}
\caption{\small Simulation parameters. The spectrum type refers to the leading non-zero order in the low wavenumber limit.  
}\label{tab:simulation_summary}
\end{table}

\begin{table}
\centering
\begin{tabular}{lrrrr}
Run Name                  & $\Rey_\lambda$ & $\ell/B$   & $\eta k_{\text{max}}$ &  Resolution \\
\\
LES\_R1                      & 76.9           & 0.17  & 1.52                   & $128^3$    \\
LES\_R2                       & 45.0           & 0.19  & 1.58                   & $64^3$\\
All others & 122.4          & 0.16 & 1.53                    & $256^3$    \\
\end{tabular}
\caption{\small Summary of the simulation parameters used in the pseudo-spectral code and the resulting IHT fields.  $\ell$ is the integral scale, $\eta$ is the Kolmogorov length scale, and $k_\text{max}$ is the maximum wavenumber. 
}\label{tab:IHT}
\end{table}

\subsection{Initial spectrum and low wavenumber limit}\label{sec:ic}

As we only expect the turbulence cloud to remain homogeneous deep within the sphere, ambiguities arise in interpreting the energy spectrum:  it can be viewed as the expectation of a random process, or merely as the Fourier transform of a deterministic function.  In order to reach the broadest conclusions possible (i.e. ones not limited to the specific initial condition), we show in Appendix \ref{sec:appendix_total_spectrum} that by invoking local homogeneity deep within the spherical region, we can estimate the total spectrum through a single realization of this flow.  The estimated spectrum approximates the true one in the limit of large $R/ \ell$, which may only be barely reached in our simulations, but in principle could be improved upon in future.  Thus we take
\begin{align}
 \widetilde{E}(\vk)
 & =\frac{1}{16 \pi^3} \int_{\R^3}  \int_{\R^3}  \vu(\vx)\cdot \vu(\vx')\,  e^{-i \vk \cdot (\vx'-\vx)} \intdiff \vx  \intdiff \vx'  \nonumber\\
 & =\frac{1}{16 \pi^3} |\mathcal{F}\{\vu\}|^2 =\frac{1}{16 \pi^3}  \frac{1}{|\vk|^2}|\mathcal{F}\{\vomega\}|^2  \nonumber\\
 & =\frac{1}{16 \pi^3} \int_{\R^3}  \int_{\R^3}  \frac{1}{|\vk|^2} \vomega(\vx)\cdot \vomega(\vx')\,  e^{-i \vk \cdot (\vx'-\vx)} \intdiff \vx  \intdiff \vx' \label{eq:Ek_vel},\\
E(k) &=\frac{1}{(2 \pi)^{2}}
    \int_{\mathbb{R}^{3}} \int_{\mathbb{R}^{3}}
    \frac{\sin \left(k\left|\vx'-\vx\right|\right)}{k\left|\vx'-\vx\right|} \vomega\left(\vx'\right) \cdot \vomega (\vx) \intdiff \vx \intdiff \vx', \label{eq:greg}
\end{align}
where we expressed the spectrum in terms of the vorticity field \citep{phillips1956final, leonard1985computing, winckelmans1993contributions, winckelmans1995some}. The spherical symmetry of the problem is used in the last step where the 3-D energy spectrum $\widetilde{E}(\vk)$ is integrated over a spherical shell to produce a scalar spectrum $E(k)$.  Expanding $E(k)$ for the low wavenumber, the odd powers vanish, giving
\begin{align}
    E(k) &= \frac{k^2}{4 \pi^2} L + \frac{k^4}{24 \pi^2} I + O(k^6) ,\label{eq:Ek_lowwavenumber}
\end{align}
where
\begin{align} \label{eq:impulse}
L&=-\frac{1}{6}  \int_{\R^3}\int_{\R^3} |\vx'-\vx|^{2} \, \vomega(\vx')\cdot\vomega(\vx)\intdiff \vx' \intdiff\vx
= \int_{\R^3} \int_{\R^3}  \vu(\vx')\cdot\vu(\vx) \intdiff \vx' \intdiff \vx
\end{align}
is the Saffman integral and
\begin{align} \label{eq:loit}
    I=\frac{1}{20}  \int_{\R^3}\int_{\R^3} |\vx'-\vx|^{4} \,\vomega(\vx')\cdot\vomega(\vx)\intdiff \vx' \intdiff\vx
=-\int_{\R^3}\int_{\R^3} |\vx'-\vx|^2 \,\vu(\vx') \cdot \vu(\vx) \intdiff \vx'\intdiff\vx
\end{align}
is the Loitsyansky integral \citep{loitsyansky1939some}.
Note that even though Eq.~(\ref{eq:Ek_vel}-\ref{eq:Ek_lowwavenumber}) are well defined for the flows with finite energy, using the velocity forms from Eq.~(\ref{eq:impulse}, \ref{eq:loit}) requires certain decay rates of the velocity field. Thus we have used the vorticity formula for the calculation of the low wavemnumber spectra.
Details of the expansion and the calculation method are given in Appendix \ref{sec:appendix_total_spectrum_B}.

The Saffman integral $L$ is related to the total momentum impulse which is an invariant of the motion and remains constant for all time. When $L \ne 0 $ the cloud will  exhibit a small-wavenumber,  $k^2$ Saffman limit \citep{saffman1967large}, whereas when $L=0$ the spectrum is of the $k^4$ Batchelor type \citep{batchelor1956large}.  The ramifications of a Saffman or Batchelor spectrum have been widely explored in IHT, but less so in other (inhomogeneous) flows.  As we discuss below, the low wavenumber spectrum can be used to derive asymptotic energy decay and integral-scale growth rates, which can be compared to those obtained for the spherical cloud.

In order to investigate this issue, we develop a procedure by which we control the value of $L$ in the initial condition.  Two factors contribute to the linear momentum. Firstly the IHT field generated from the pseudo-spectral code is continuously divergence-free but not necessarily discrete divergence-free as required by the finite volume FLGF scheme. Secondly, the windowing process will introduce extra non-solenoidality. i.e., given a divergence-free velocity field $\boldsymbol{u}$ and a scalar window function $\Phi(r)$, $\Delta \cdot (\Phi(r) \boldsymbol{u}(\vx))\neq 0$ in general. Both of these non-solenoidal components are projected out at the very first time step and this projection will introduce an impulse. The result of this impulse as mentioned in \citet{batchelor1967introduction} is a $1/|\vx|^3$  decaying velocity field of the form
\begin{align} \label{eq:helmholtz}
    \lim_{x\rightarrow \infty}\mathbf{u}(\vx) =
    \frac{1}{8 \pi} \nabla \left[
        \nabla \left(\frac{1}{|\vx|}\right) \cdot \int_{\mathbb{R}^3} \vx' \times \vomega(\vx')
   \intdiff \vx'\right]
    .
\end{align}
This suggests a way to cancel the impulse in order to have a $k^4$ type spectrum: one can add a vortex ring with an opposite impulse to the initial velocity field. More specifically, we add a Stokes vortex ring with velocity \citep{kambe1975generation, cantwell1986viscous}
\begin{align}\label{eq:cancel}
\vu(r, \theta) = \frac{1}{(2\pi)^{3/2}\zeta^3}\left[W\left(\frac{r}{\zeta};1 \right) \boldsymbol{\gamma} - W\left(\frac{r}{\zeta};3 \right) \gamma \cos \theta \, \hat{\mathbf{e}}_{r}  \right],
\end{align}
where $\boldsymbol{\gamma}$ is the impulse of the vortex ring, $\theta$ is the angle between $\boldsymbol{\gamma}$ and the unit vector $\mathbf{e}_{r}$, $\zeta$ controls the size of the ring and
\begin{align}
    W(\rho;b)
    &=e^{-\rho^2/2} -  \frac{b}{\rho^3} \left[ \sqrt{\frac{\pi}{2}}\operatorname{erf}\left(\frac{\rho}{\sqrt{2}}\right)-\rho e^{-\rho^{2} / 2}\right]
    .
\end{align}
We chose $\zeta/R = 0.19$ and performed an LES computation with this cancellation, referred to as case LES\_IC2 in table \ref{tab:simulation_summary}. Except for LES\_IC2, all other cases are conducted without the cancellation.

While our method of manipulating the initial condition in order to cancel the finite impulse is arbitrary, it is effective in the sense that the added vortex ring quickly interacts with the turbulence leading.  This was verified by monitoring the difference between simulations initialized with and without the cancellation. The results showed that in less than one initial large-eddy turnover time, the difference field was decorrelated with the added vortex ring.  Thus we conclude that the two simulations can be regarded as representing (different random realizations of) locally homogeneous turbulence that differ significantly only in their low wavenumber spectrum.

\begin{figure}
    \centering
    \begin{overpic}[width=.75\textwidth]{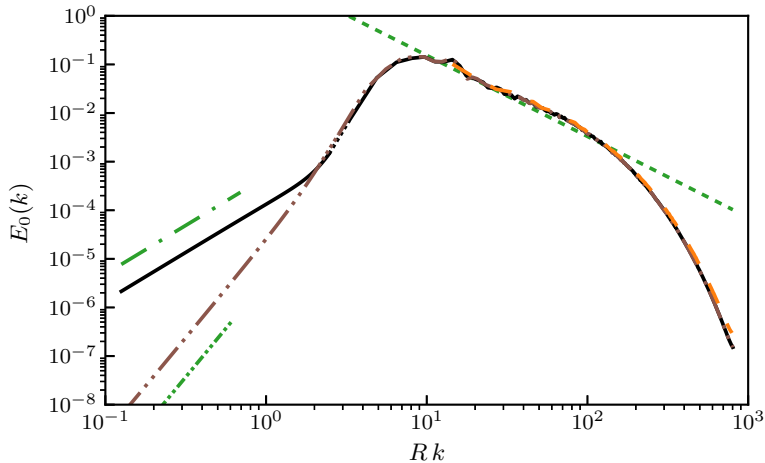}
    \end{overpic}
    \caption{ \small Energy spectrum of
    (1) the initial condition of simulation DNS\_0. (\protect\blackline), where the low wavenumber limit (left of the dotted region) is calculated through an expansion method (Appendix \ref{sec:appendix_total_spectrum_B});
    (2) DNS\_0 with the initial impulse cancelled using equation~\eqref{eq:cancel} (\protect\browndashdotdot);
    (3) the corresponding IHT field multiplied by the ratio between the spherical region volume $\frac{4}{3} \pi R^3$ and the cubic domain volume $B^3$ (\protect\orangedash);
    (4) a guide line for $1.6 \varepsilon^{2/3} k^{-5/3}$ scaled with the same ratio, where $\varepsilon$ is the dissipation rate in the original IHT field (\protect\greendot);
    (4) a slope of $k^2$ (\protect \greendotdash) and $k^4$ (\protect \greendensedashdotdot) for the low wavenumber limit.
    }\label{fig:intial_Ek}
\end{figure}

Figure \ref{fig:intial_Ek} shows the resulting initial energy spectrum $E_0(k)$ of a spherical region of turbulence field corresponding to the condition of DNS\_0, superposed on another spectrum where equation~\eqref{eq:cancel} was used to cancel the impulse (i.e. the initial condition, after filter, for the case `LES\_IC2' in table \ref{tab:simulation_summary}). Also plotted is the energy spectrum of the original IHT field scaled by the ratio between the volume of the sphere and the cubic domain size $B^3$.  We see that the $k^{-{5/ 3}}$ portion of the spectrum from IHT is retained in the spherical cloud, whereas the low wavenumber behavior is controlled by the resulting impulse (or its absence).

\subsection{Resolution}\label{sec:resolution}

\begin{figure}
    \centering
    \begin{overpic}[width=.75\textwidth]{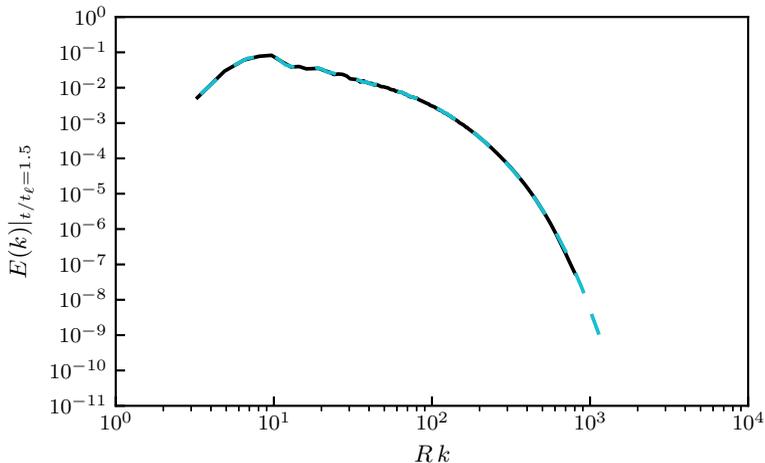}
    \end{overpic}
    \caption{ \small Energy spectrum of
    (1) DNS\_0 (\protect\blackline) and
    (2) a DNS calculation at $3/2$ times the resolution as that in the case DNS\_0 (\protect\cyandash), at $t/t_\ell=1.3$.
    }\label{fig:res_check}
\end{figure}

For DNS\_0, the IHT field used to generate the initial condition has $\Rey_\lambda=122.4$ and uses a computational domain of $256^3$ in the pseudo-spectral code with $\eta k_{\text{max}}>1.5$ to ensure that it is fully resolved. The same resolution (same number of points used for every length scale $B$) is used in the LGF solver for the turbulence cloud. To guarantee this resolution is also sufficient for the finite volume solver, another DNS simulation of $3/2$ times the resolution is performed up to $1.3$ initial large eddy turnover time. The difference in the total kinetic energy is about $0.23\%$ and the maximum relative difference in the spectra for all wavenumber $kR$ is about $1\%$ which is shown in figure \ref{fig:res_check}.

\section{Qualitative evolution}
\label{sec:viz}
\subsection{DNS}\label{sec:DNS}
\begin{figure}
    \centering
    \begin{overpic}[width=.79\textwidth]{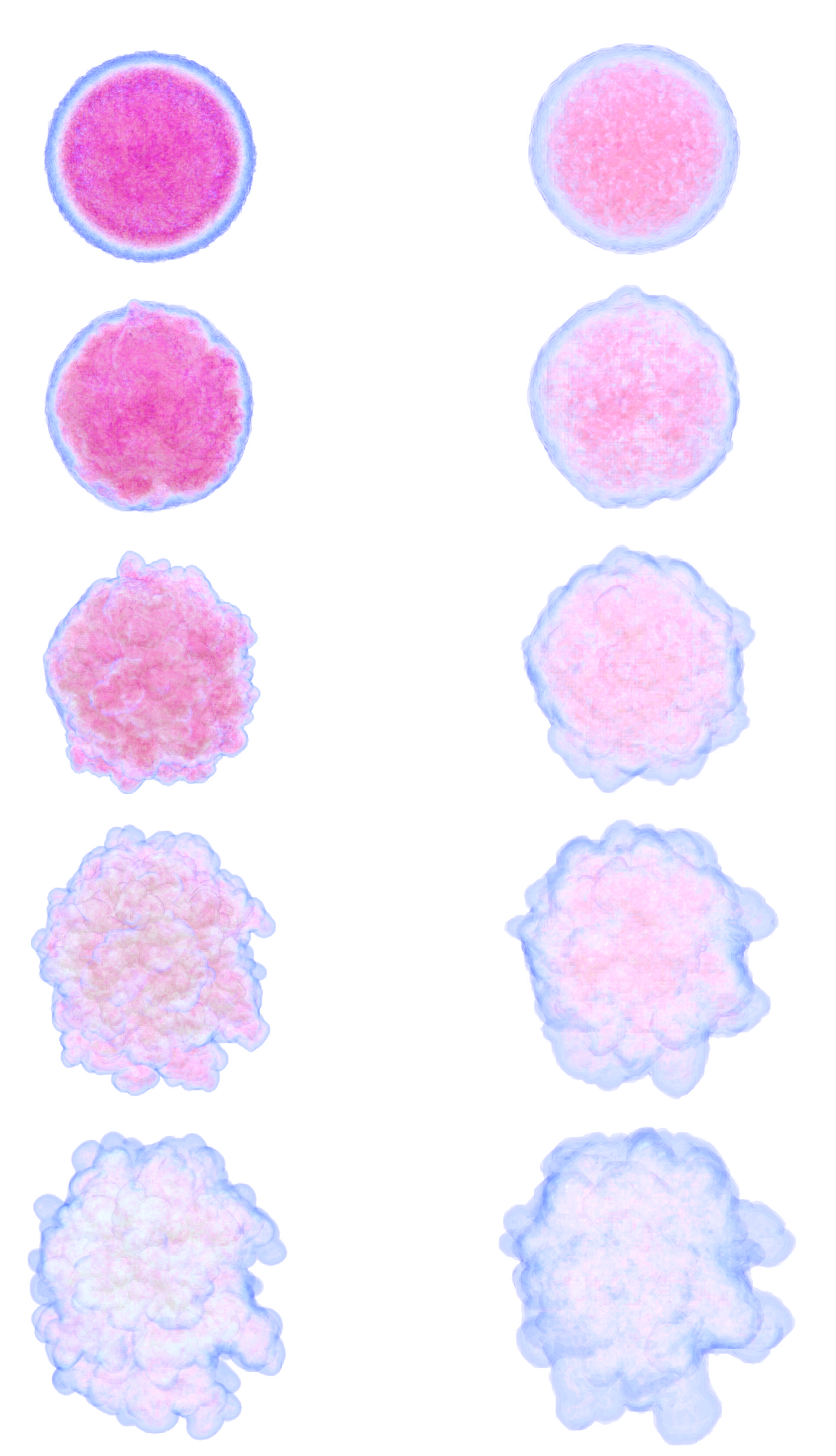}
        \put(-1,  97)   {\small(a)}
        \put(30,  97)   {\small(b)}
    \end{overpic}
     \caption{\small Vorticity magnitude iso-surface of (a) DNS\_0 and (b) LES\_0 at $t/t_\ell=0, 1.72, 4.02, 8.60,  17.47$ from top to bottom, where $t_\ell$ is the large eddy turnover time of the initial IHT field. }\label{fig:DNS}
\end{figure}

First we perform DNS of the spherical cloud of turbulence corresponding to the case `DNS\_0' in table \ref{tab:simulation_summary}.
The flow evolution is shown in figure \ref{fig:DNS}a. Instantaneous vorticity magnitude iso-surfaces
at $t/t_{\ell}=0,1.7, 4.0, 8.6,  17.5$ are given, where $t_\ell$ is the large eddy turnover time of the original IHT field. The iso-surface of the lowest vorticity magnitude represents the TNTI.
This interface is sufficiently thin \citep{mathew2002some} that using a lower minimum vorticity magnitude would not affect the boundary envelope noticeably.
At $t/t_{\ell}=0$ the turbulence is contained within a spherical region defined by the window function.
As the turbulence evolves, the transition region is mixed with the turbulence inside and becomes gradually indistinguishable around $t/t_{\ell}\sim 1.5$.  At $t/t_{\ell}\sim 4.0$ more fine features have developed near the boundary while the general spherical shape is still maintained. Around $t/t_{\ell}\sim 8.6$, small features start to merge and create protrusions. Meanwhile the general shape has also become more ellipsoidal.  The DNS flow evolution is simulated up to $t/t_{\ell}=17.5$. From $t/t_{\ell}=8.6$ to $17.5$ the cloud of turbulence becomes more irregular, and finer scales are less evident as the turbulence decays.

\subsection{Comparison between DNS and LES}\label{sec:DNSLES_visualization}
DNS is applied to study the more active early stage evolution of a turbulence cloud but it is computationally expensive to reach late times. To study the long-term behavior we turn to LES calculations of the same setup.  To ensure LES calculations are able to accurately capture the abiding features of the flow, we qualitatively compare the evolution for DNS and LES of the same case.

The initial condition for the LES run is created in the following way: first, the same IHT field from case DNS\_0 is  spectrally filtered from $256^3$ to $32^3$, keeping only $(1/8)^3$ of its original spectrum; second, the same recipe (tiling and spherical windowing) is used with the filtered turbulence field to create a spherical region of under-resolved turbulence. This field is then given to the LGF finite-volume solver with the SGS model turned on. This simulation corresponds to the case `LES\_0' in table \ref{tab:simulation_summary}.

Figure \ref{fig:DNS} also compares DNS\_0 (\ref{fig:DNS}a) and LES\_0 (\ref{fig:DNS}b) at $t/t_\ell=0,  1.72, 4.02, 8.60,  17.47$. To ensure that the difference in grid resolution between DNS and LES would not affect the visualization, all iso-surfaces are re-sampled to the same grid. The LES captures the general shape and most of the large-scale features such as the radius, the ellipticity, the sizes and locations of the protrusions. On the other hand, some small-scale features near the boundary are missed. We also noticed that the vorticity is less intense in the LES run (the crimson regions) especially towards the early stage.  All of these differences are to be expected, as LES is designed to capture the statistical properties of the turbulence (and specifically their influence on the largest scales).  Nevertheless, over the time range displayed in figure \ref{fig:DNS} there is little decorrelation of the large scales in DNS and LES originating from the same initial condition.  To further quantify the comparison, in \S~\ref{sec:statistics}, four statistical measures are introduced and applied to both cases DNS\_0 and LES\_0.

\section{Quantitative evolution}\label{sec:statistics}
\subsection{Statistical measures for DNS and LES}\label{sec:stats_DNSLES}

In this section we quantify the initial evolution of the cloud of turbulence using statistical measures. DNS results are compared with LES during the initial decay period up to about $t / t_\ell \simeq 20$.

\begin{figure}
     \centering
         \includegraphics[width=0.75\textwidth]{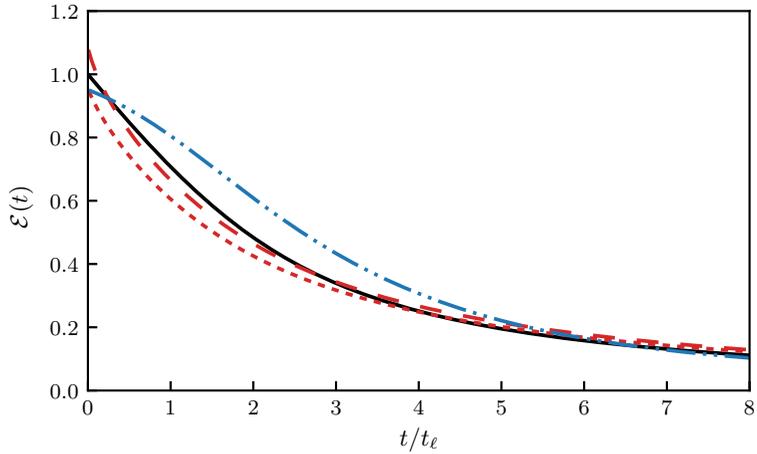}
         \caption{\small Decay of the kinetic energy ${\cal E}(t)$ for different simulations: DNS\_0 (\protect\blackline);  LES\_0 resolved kinetic energy (\protect\reddot); LES\_0 total kinetic energy (\protect\reddash); an under-resolved DNS (\protect \bluedotdash). The initial kinetic energy in DNS\_0 is used to normalize all simulations.}
        \label{fig:Ek_decay}
\end{figure}

Firstly the kinetic energy ${\cal E}(t)$ decay is studied. Results from three simulations are compared in figure \ref{fig:Ek_decay}: (1) DNS\_0; (2) LES\_0 and (3) an under-resolved DNS (the same setup as LES\_0 but with SGS model turned off).
DNS\_0 should be regarded as the most accurate case among all three and its value at $t=0$ is used to normalize all results.
For LES\_0 we show both the kinetic energy resolved by the grid and a `total kinetic energy' which is the sum of the resolved energy and the estimated subgrid energy predicted by the SGS model. The total kinetic energy in LES\_0 compares well with DNS\_0.  The initial resolved energy in LES\_0 is smaller than that in DNS\_0 owing to the spectral filtering process discussed in \S~\ref{sec:initial_conditions}.
On the other hand the under-resolved DNS shows evident energy pile-up due to the lack of the SGS model.
After $t/t_\ell \simeq 8$, the resolution of case LES\_0 is high enough to resolve all flow scales due to the decay and it is effectively `DNS' after this point.

\begin{figure}
    \centering
    \begin{overpic}[width=.75\textwidth]{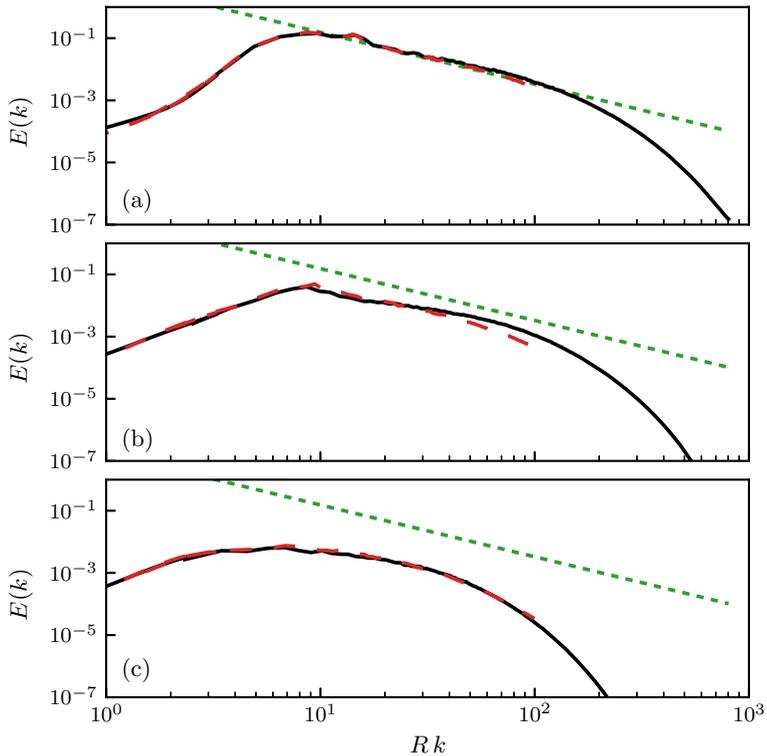}
        \put(15,73)   {\small(a)}
        \put(15,42)   {\small(b)}
        \put(15,12)   {\small(c)}
    \end{overpic}
     \caption{\small Total energy spectrum at (a) $t/t_\ell=0$, (b) $t/t_\ell=4.02$ and (c) $t/t_\ell=17.47$ for DNS\_0 (\protect\blackline) and LES\_0 (\protect\reddash). The same guide line for $k^{-{5 / 3}}$ as in figure \ref{fig:intial_Ek} is also given (\protect\greendot).}
    \label{fig:total_spectrum}
\end{figure}

Secondly the total energy spectrum defined in \S~\ref{sec:ic} (equation~\eqref{eq:greg}) is applied again here. Results for DNS\_0  and LES\_0 are compared in figure \ref{fig:total_spectrum} at three time instants $t/t_\ell=0,4.02,17.47$, corresponding to the visualizations in figure \ref{fig:DNS}. The agreement is significant and LES seems to only suppress the high wavenumbers slightly.   For both cases, the initial condition features about a decade of inertial-range turbulence with a $k^{-{5 / 3}}$ spectrum; as expected, this region shrinks (from the high wavenumbers) as the turbulence decays.

The third statistical measure we consider is the energy spectrum on a spherical shell.  The flow is designed to be homogeneous in the azimuthal and polar directions but is only so in the radial direction deep within the cloud. As shown in \S~\ref{sec:DNS} visually the cloud also undergoes radial growth over time. This non-uniformity in the radial direction suggests that one should further characterize the energy spectrum as a function of the radius $r$ and time $t$.  A special spectrum defined on a spherical shell of a given radius is applied \citep{Lombardini2014}. Similar to the total energy spectrum $E(k)$, it seeks a relationship between the energy and the wavenumber, where the wavenumber on a spherical shell is defined using the spherical harmonics. The spherical shell wavenumber and the classical wavenumber defined using Fourier transform can be related via the Laplace operator. This relation also connects the spherical shell spectrum to the classical energy spectrum.

Following the detailed derivation given by \citet{Lombardini2014},  we acquire the shell spectrum for a field $f_r(\theta, \phi)$ defined for a given raidus $r$ by expanding the field using spherical harmonics:
\begin{align}\label{eq:harmonicsExpansion}
    f_r(\theta, \phi)=\sum_{\ell=0}^{\infty} \sum_{m=-\ell}^{\ell} f_{\ell m} Y_{\ell m}(\theta, \phi),
\end{align}
where
\begin{align}
    Y_{\ell m}(\theta, \phi)=\left\{\begin{array}{ll}{N_{(\ell, m)} P_{\ell}^{m}(\cos \theta) \cos (m \phi)} & {m \geqslant 0} \\ {N_{(\ell,|m|)} P_{\ell}^{m |}(\cos \theta) \sin (|m| \phi)} & {m<0}\end{array}\right. ,
\end{align}
with $P_{\ell}^{m}$ being the associated Legendre polynomials, $N_{(\ell, m)}$ being the normalization constant and $\ell$ being the equivalent wavenumber. The wavenumber $\ell$ is then  related to the classic wavenumber $k$ defined through the Fourier transform by,
\begin{align}
    k^{2}=\ell(\ell+1) / r^{2}.
\end{align}
Assuming a power law for the energy spectrum $E(k) \sim k^{-\alpha}$ one has the following relationship between the energy spectrum and the shell spectrum,
\begin{align}\label{eq:Cl}
     E(k)\sim k^{-\alpha} \sim\ell C_{\ell} ,
\end{align}
where
\begin{align}
    C_{\ell}=\frac{1}{2 \ell+1} \sum_{m=-\ell}^{\ell} f_{\ell m}^{2}.
\end{align}
Equation \eqref{eq:Cl} suggests that one can understand the shell spectrum $C_\ell$ in a similar way as the classical energy spectrum $E(k)$.

\begin{figure}
    \centering
    \begin{overpic}[width=.95\textwidth]{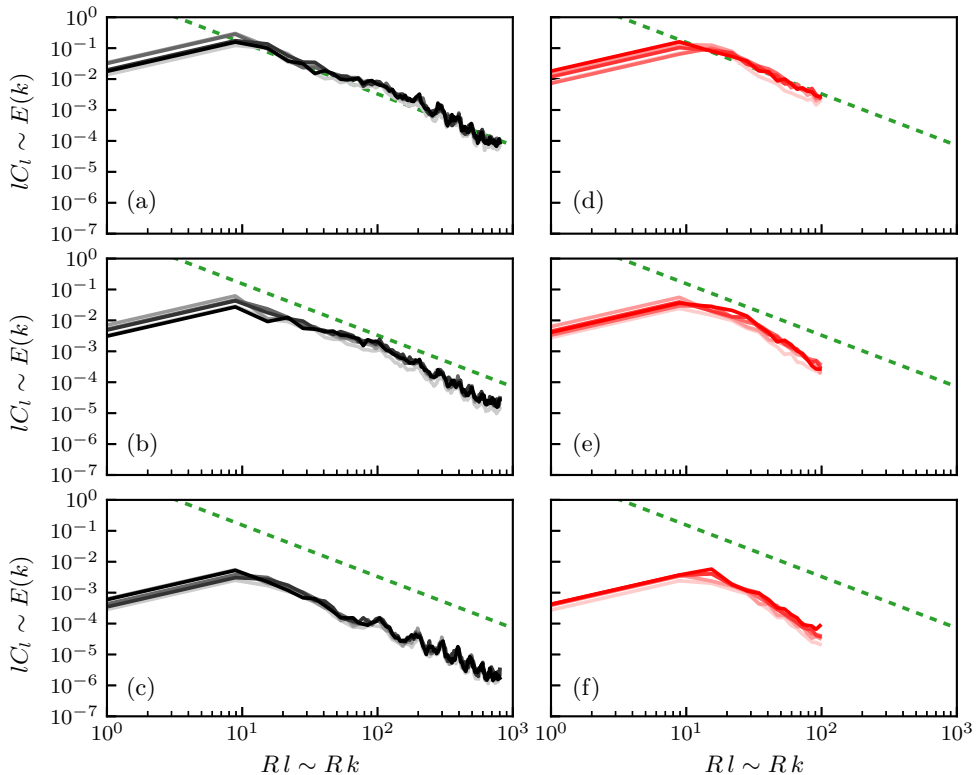}
        \put(12,59)   {\small(a)}
        \put(12,34)   {\small(b)}
        \put(12,9)    {\small(c)}
        \put(58,59)   {\small(d)}
        \put(58,34)   {\small(e)}
        \put(58,9)    {\small(f)}
    \end{overpic}
     \caption{\small
     Spherical-shell spectrum at different radii and times.
     The left column (a-c) are results from DNS\_0 at $t/t_\ell=0, 4.02, 17.47$ respectively. The right column (d-f) are results from LES\_0 at the same times.  For each figure, the gradation in color corresponds to radii $r/B=(0.32,0.48,0.64,0.80,0.96)$ from darkest to lightest shade. An equivalent guide line for $k^{-{5 / 3}}$ as in figure \ref{fig:intial_Ek} is given (\protect\greendot).
    }
     \label{fig:sphericalShell}
\end{figure}

The shell spectra for DNS\_0 and LES\_0 are shown in figure \ref{fig:sphericalShell}. Results for various radii ($r/B=0.32,0.48,0.64,0.80,0.96$) and time instants ($t/t_\ell=0, 4.02, 17.47$) are given. At $t/t_\ell=0$ (figure \ref{fig:sphericalShell}: a, d) all 5 curves collapse together as expected since they represent the original IHT field. The energy decays over time, but the dependence of the shell spectrum on the radius $r$ is weak. It seems that the boundary does not have a strong effect on the turbulence decay.  This evidence further supports the assumption of local homogeneity that underpins our definition of the total spectra used above, as discussed in the appendix, at least up through the times considered here.

\subsection{Long-term statistics and low wavenumber behavior}\label{sec:longterm_statistics}
\begin{figure}
    \centering
    \begin{overpic}[width=0.75\textwidth]{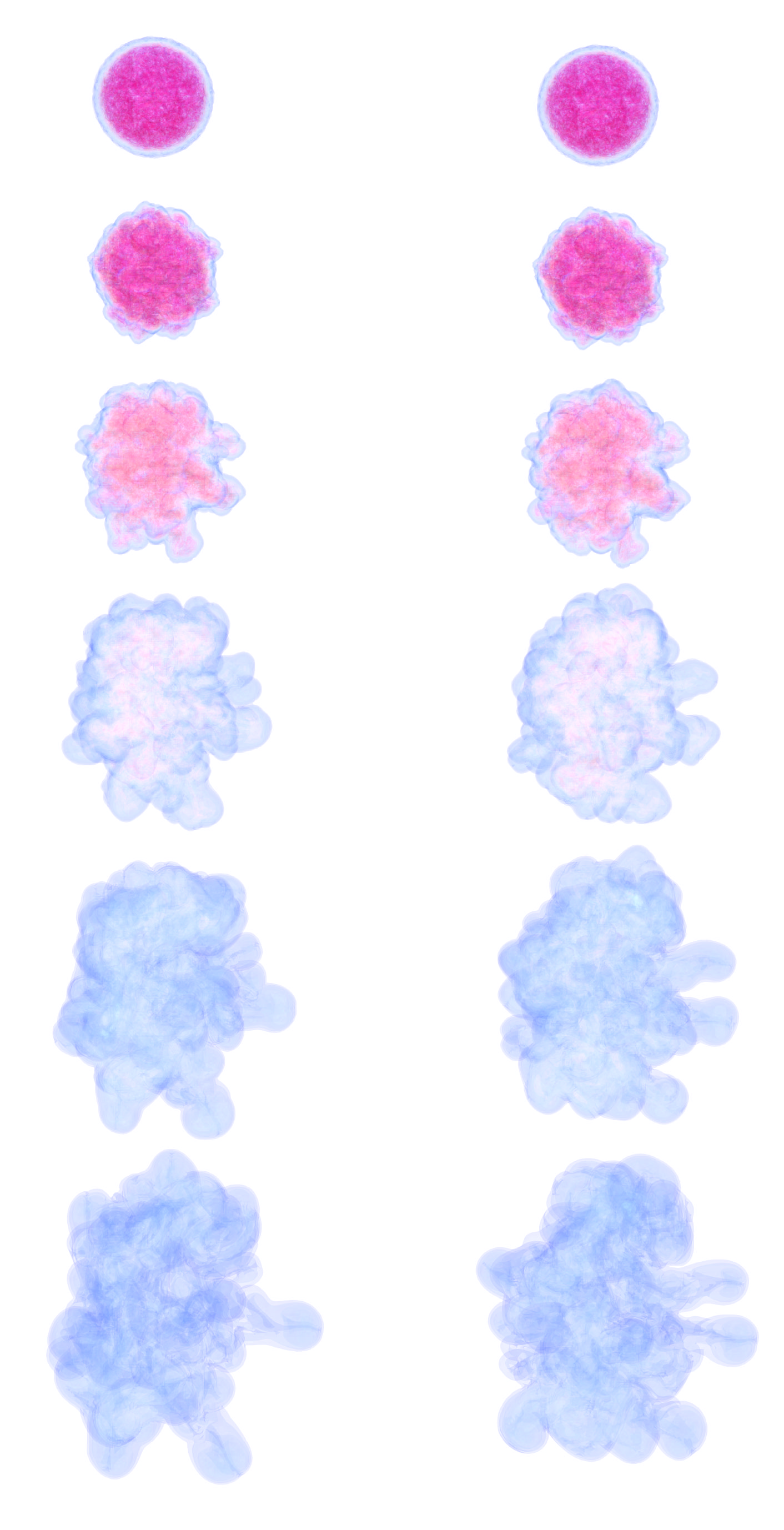}
    \put(0,98)   {\small(a)}
    \put(28,98)   {\small(b)}
    \end{overpic}
    \caption{\small Long-term evolution of (a) LES\_0 and (b) LES\_IC2 at $t/t_\ell=0, 4, 17, 66, 143, 263$ from top to bottom. }\label{fig:longterm_LES}
\end{figure}

The long-term evolution of a turbulence cloud is studied through LES. Two cases LES\_0 and LES\_IC2 from table \ref{tab:simulation_summary} are simulated up to $t/t_\ell=400$ where LES\_0 has a $k^2$-type initial spectrum while LES\_IC2 has a $k^4$-type. The evolution is visualized in figure \ref{fig:longterm_LES}. The spread of the cloud is similar in both cases, but details of the large-scale structures are different.

\begin{figure}
    \centering
    \begin{overpic}[width=0.75\textwidth]{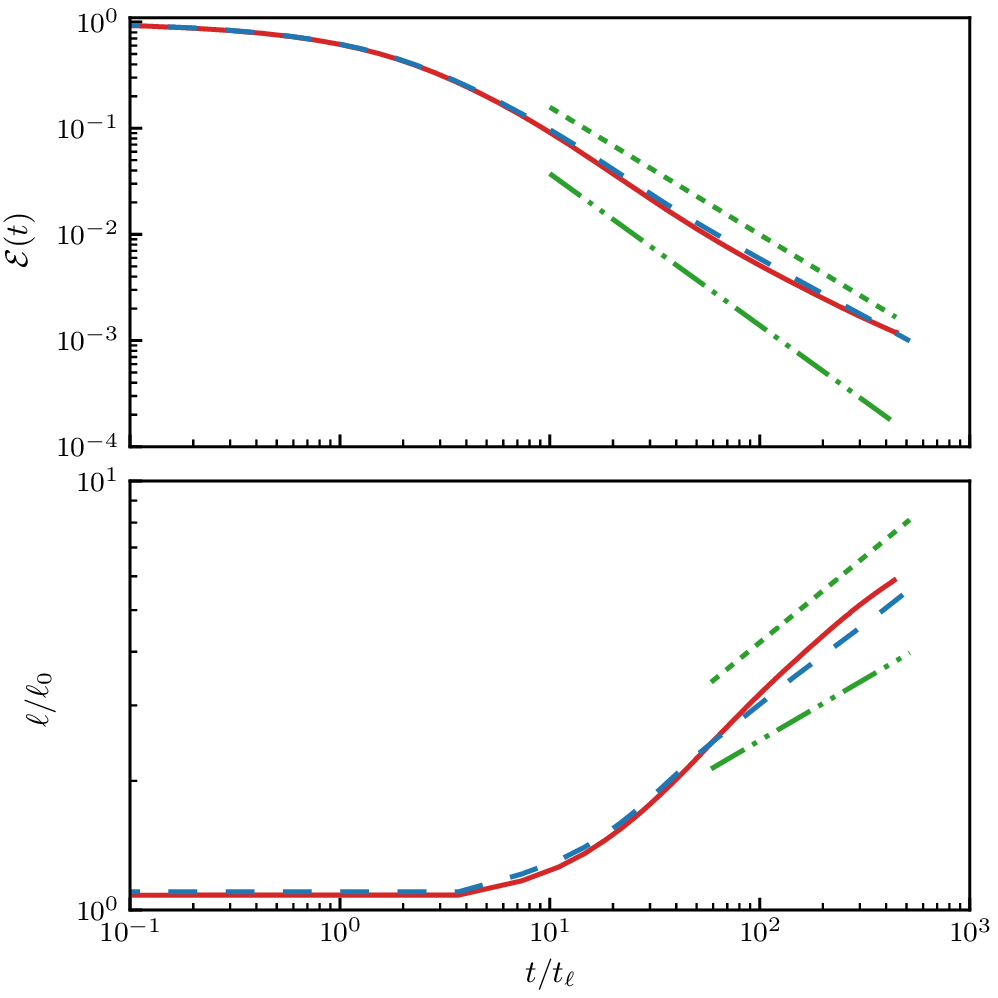}
        \put(18,62)   {\small(a)}
        \put(18,45)   {\small(b)}
    \end{overpic}
    \caption{ \small Long-term evolution of (a) the kinetic energy decay compared with asymptotic behavior of Saffman IHT $\mathcal{E}(t) \sim t^{-6/5}$ (\protect\greendot) and Bathelor IHT $\mathcal{E}(t) \sim t^{-10/7}$ (\protect\greendashdotdot);  (b) the integral scale growth for case LES\_0 (\protect \redline) and LES\_IC2 (\protect\bluedash) up to $t/t_\ell = 400$ compared with asymptotic behavior of Saffman IHT $\ell \sim t^{2/5}$ (\protect\greendot) and Bathelor $\ell \sim t^{2/7}$ (\protect\greendashdotdot). }\label{fig:longterm_Ek}
\end{figure}

Figure \ref{fig:longterm_Ek}a shows the long-term evolution of the kinetic energy decay. In the case of IHT, \citet{saffman1967large} predicts an asymptotic decay rate of $t^{-6/5}$ for the $k^2$ turbulence and \citet{kolmogorov1962refinement} predicts a decay rate of $t^{-10/7}$ for the $k^4$ spectrum (both guide lines are indicated in the figure). However in both the case of LES\_0 and LES\_IC2, the decay is similar and closer to the Saffman scaling. Figure \ref{fig:longterm_Ek}b shows the evolution of the integral scale over time for LES\_0 and LES\_IC2, compared to the theoretical asymptotic growth rate for Saffman IHT ($t^{2/5}$) and Batchelor IHT ($t^{2/7}$). As in the energy decay, both cases are closer to the Saffman type.

\begin{figure}
    \centering
    \begin{overpic}[width=0.75\textwidth]{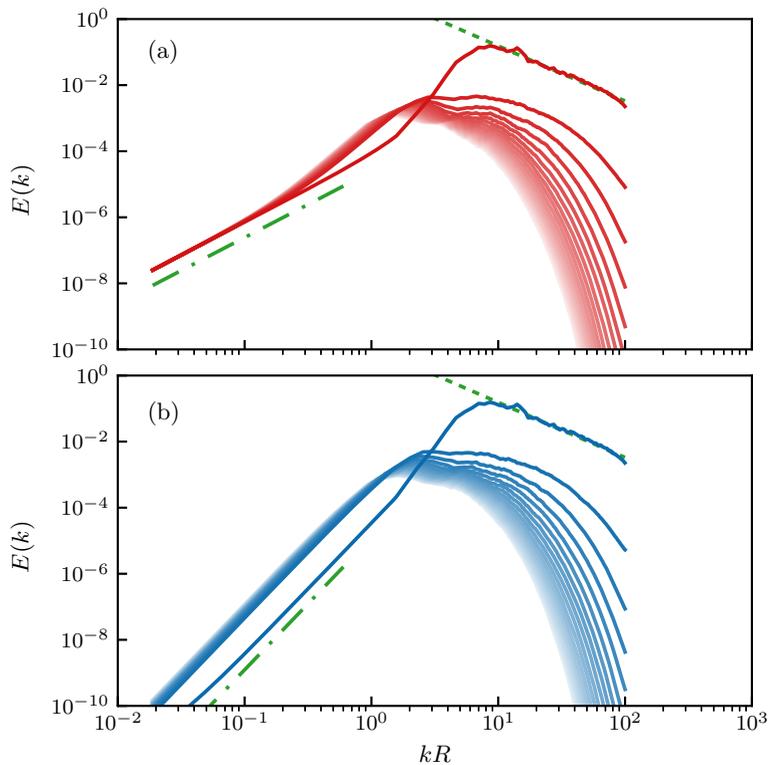}
        \put(18,93)   {\small(a)}
        \put(18,46)   {\small(b)}
    \end{overpic}
   \caption{ \small Long-term evolution of the total spectrum for (a) LES\_0 and (\protect \redline) (b) LES\_IC2 (\protect \blueline), up to $t/t_\ell=500$ with $\Delta t/t_\ell=15$ between each line. In figure (a) guide lines for $k^{-5/3}$ (\protect\greendot) and $k^2$ (\protect\greendotdash) are given.  In figure (b) guide lines for $k^{-5/3}$ (\protect\greendot) and $k^4$ (\protect\greendotdash) are given.  }\label{fig:longterm-spectrum}
\end{figure}

\begin{figure}
    \centering
    \begin{overpic}[width=0.75\textwidth]{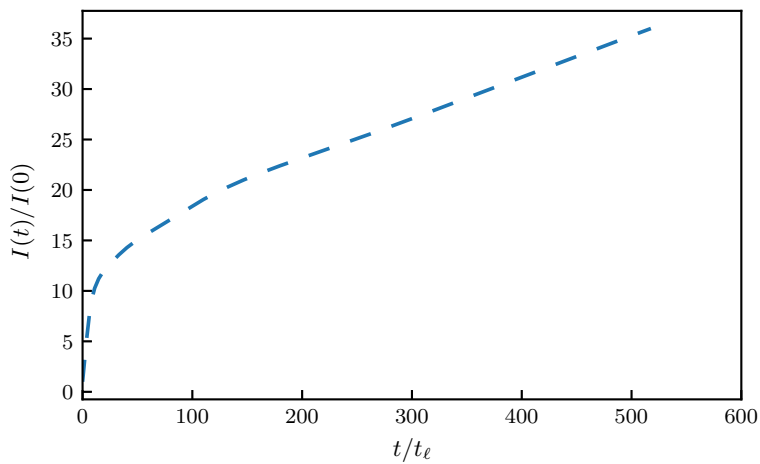}
    \end{overpic}
   \caption{ \small
   Long-term evolution of the normalized Loitsyansky integral $I(t)/I(0)$ of LES\_IC2 ($k^4$ type), up to $t/t_\ell=500$. }\label{fig:loitsyansky}
\end{figure}

This apparent discrepancy with the theory can be clarified by examining the long-term decay of the total energy spectrum depicted in figure \ref{fig:longterm-spectrum}, which shows results for both LES\_0 ($k^2$) and LES\_IC2 ($k^4$) cases. The $k^4$ spectrum is similar to that reported in \citet{ishida2006decay}.  As expected the coefficient of the limiting $k^2$ spectrum for LES\_0 is invariant, but the coefficient of the $k^4$ term for LES\_IC2 is increasing over time first rapidly up to about $t / t_{\ell} \simeq 20$, and then more slowly, as shown in figure~\ref{fig:loitsyansky}.  The coefficient of the $k^4$ term is proportional to the Loitsyansky integral $I$ given by equation~\eqref{eq:loit}  which is assumed constant in the theory under the assumption that remote points be statistically independent \citep{loitsyansky1939some}.

\begin{figure}
    \centering
    \begin{overpic}[width=0.9\textwidth]{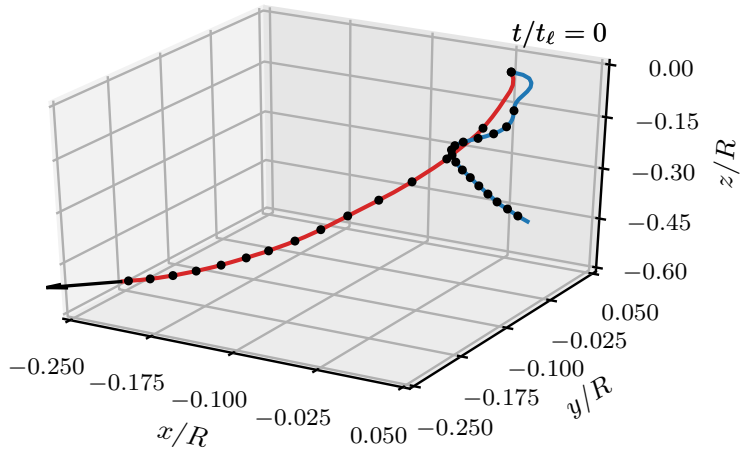}
    \end{overpic}
    \caption{ \small Trajectory of the center of the turbulence cloud from $t/t_\ell=0$ to $400$ for LES\_0 (\protect\redline) and LES\_IC2 (\protect\blueline)  with every marker sepearted by $\Delta t/t_\ell=30$. For LES\_0 the direction of the initial impulse is indicated with the arrow at the end of the trajectory. For LES\_IC2 the impulse is zero.}\label{fig:meander}
\end{figure}

Superposition of the energy spectra for cases LES\_0 and LES\_IC2 shows they are similar for $k R > 1$, corresponding to a wavelength $\lambda = 2 \pi R$.  Apart from the very largest scales, which cannot be seen in visualizations like figure~\ref{fig:longterm_LES}, the two simulations are otherwise statistically similar.  The weak vortex ring and vortex ring dipole associated with the $k^2$ and $k^4$ terms would only become evident as $t \rightarrow \infty$, after which all turbulence will decay.  However at the same time, the properties are entirely predictable from their initial conditions.  Indeed, figure~\ref{fig:meander} shows that the entire cloud of LES\_0 meanders in space but eventually attains a trajectory that is associated with the initial impulse.

The predictability of the long-term evolution from the initial condition argues against any universality of the very largest scales of the spherical cloud of turbulence.  While we expect the wavenumber spectrum for $k R > 1$ is approximately universal, the low wavenumber behavior is always an artifact of initial and boundary conditions.

While this lack of universality is perhaps unsurprising, the veracity of the Saffman-type decay-rate predictions even in the absence of a $k^2$ spectrum is interesting.  Consider the process by which the initial $k^4$ spectrum is created for the case LES\_IC2, whereby a weak vortex ring is added to offset the initial impulse associated with windowing the IHT field.  While we superposed this ring at the center of our cloud, we could have cancelled the impulse by adding a ring at any position, even one very far from the cloud.  Over the timescale simulated here, the results would be identical to those of the $k^2$ cloud, and one would have to go to even lower values of $k R$ in order to see the ultimate $k^4$ behavior.

\begin{figure}
    \centering
    \begin{overpic}[width=0.75\textwidth]{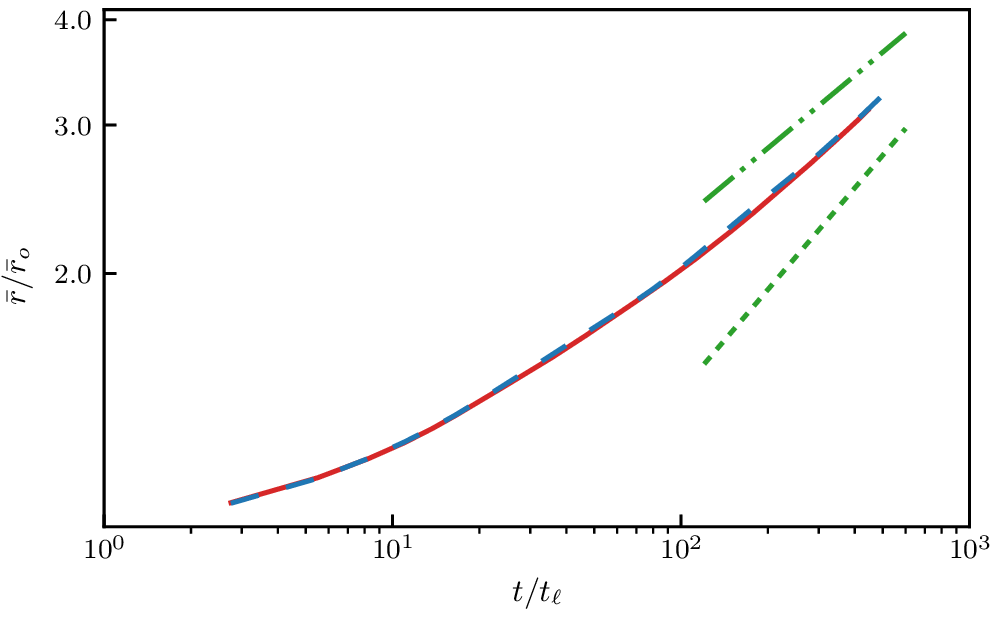}
    \end{overpic}
    \caption{ \small Comparison of the long-term mean radius $\overline{r}$ for $p=2$ between LES\_0 (\protect \redline) and LES\_IC2 (\protect\bluedash) up to $t/t_\ell \sim 400$. Results are normalized with their initial mean radii $\bar{r}_o$. Guide lines for $t^{2/5}$ (\protect\greendot) and $t^{2/7}$ (\protect\greendashdotdot) are also given.}\label{fig:longterm_radius}
\end{figure}

Lastly we consider the radial growth of the turbulence cloud over time. Because the cloud does not hold its sphericity we define the radius by a statistical moment
\begin{align}
\overline{r} &=\left(\frac{\int u^{2}|\vx -\vx_c|^{p} \intdiff \vx }{\int u^{2} \intdiff \vx }\right)^{1 / p},
\end{align}
where $u$ is the velocity magnitude and $\mathbf{x_c}$ is the center of the turbulence cloud, defined using
\begin{align}
\mathbf{x_c}=\frac{\int \mathbf{x}\, u^2 \intdiff \vx}{\int u^2 \intdiff \vx}.
\end{align}
A definition of the center is necessary because, as discussed in \S~\ref{sec:initial_conditions}, the final stage of a turbulence cloud is a large vortex ring drifting in the direction of the impulse.
Also $p \leq 2$ is needed for $\overline{r} $ to exist as the velocity field $\vu(\mathbf{x}) \sim 1/|\mathbf{x}|^3$ as $|\mathbf{x}|\rightarrow \infty$. Here we only consider the case  when $p = 2$ for simplicity.  The results  between LES\_0 and LES\_IC2 are shown in figure \ref{fig:longterm_radius}. The mean radius growth in time is almost the same for both cases, and approaches a power-law behavior that is similar to the growth of the integral scale.

\section{Vortex ring ejections}\label{sec:vortex_rings}

One of the most distinctive features in the late-stage evolution of a turbulence cloud (figure \ref{fig:longterm_LES}) is that it ejects vortex rings of roughly the same size from its boundary. In this section, we investigate the relation between the size of the vortex rings and properties associated with the turbulence.

\begin{figure}
    \centering
    \begin{overpic}[width=.95\textwidth]{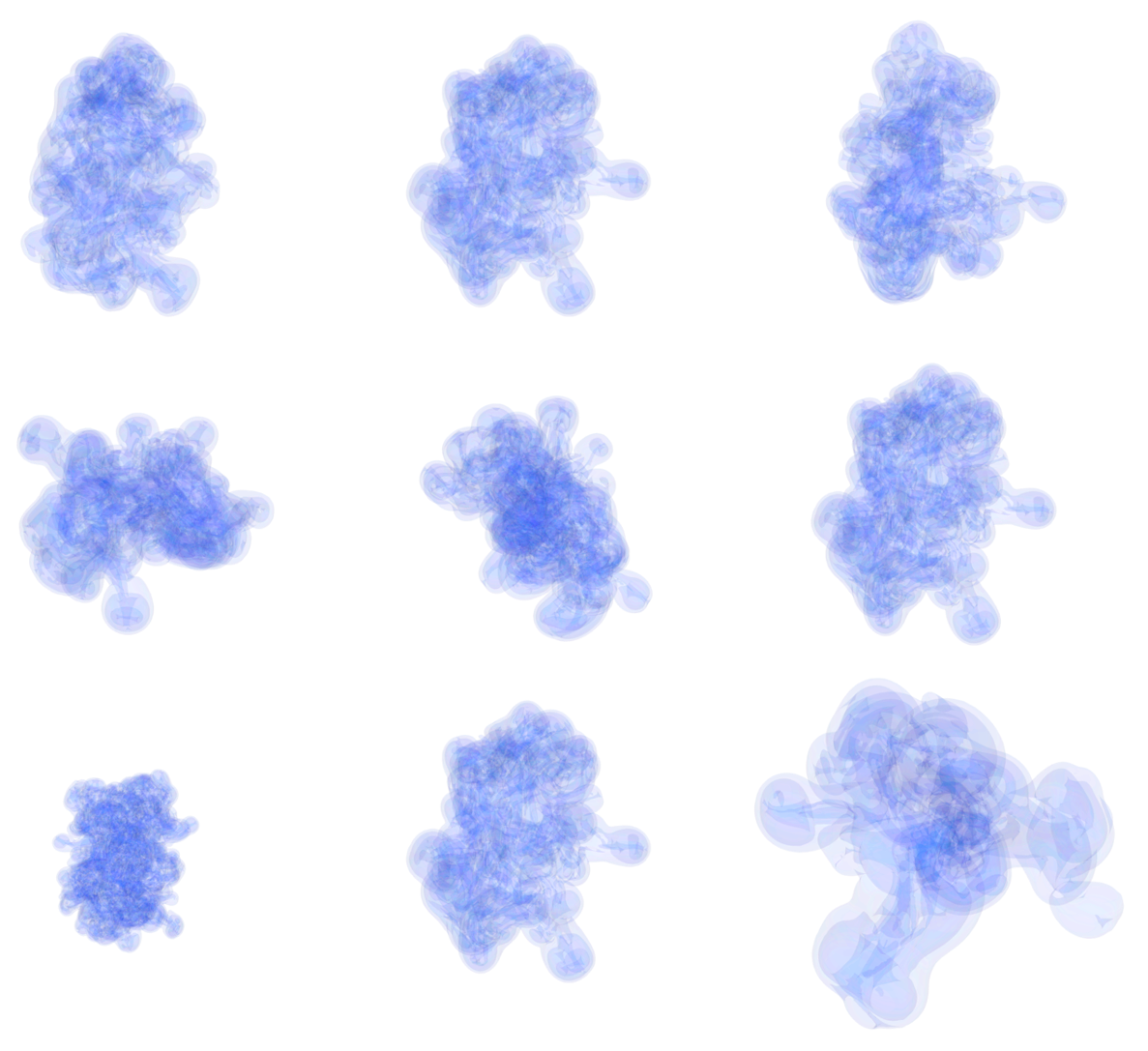}
        \put(-2, 91) {\small $\sigma/R=[0.05,0.1,0.2]$}
        \put(-2, 59) {\small $\Rey_\lambda= [45.0, 76.9, 122.4]$}
        \put(-2, 30) {\small $B/R=[0.05,0.1,0.2]$}
    \end{overpic}
     \caption{\small Long-term turbulence cloud evolution with vortex ring ejections for cases defined in table \ref{tab:simulation_summary}. First row: LES\_D1, LES\_0, LES\_D2; second row: LES\_R2, LES\_R1, LES\_0; third row: LES\_B1, LES\_0, LES\_B2.
     }\label{fig:ejectionComparison}
\end{figure}

We create LES simulations which independently vary the three independent nondimensional parameters that control the initial conditions.  Long-term evolution for all three pairs at $t=260$ are provided in Figure \ref{fig:ejectionComparison}.  The first parameter is the width of the transition region associated with the windowing function, $\sigma/R$, which is varied from [0.05,0.1,0.2] in three cases [LES\_D1, LES\_0, LES\_D2].  We see that the width of the transition region has little influence on the number or scale of the ejections.  Next, we consider varying the microscale Reynolds number $\Rey_\lambda$, which is varied from [45.0, 76.9, 122.4] in three cases [LES\_R2, LES\_R1, LES\_0].  Again, though each cloud has a different range of scales present, the vortex ejections occur again at roughly the same scale.  Finally, we vary initial integral scale $\ell/R$, by changing the size of the initial periodic box to the sphere radius, $B/R$ over the range [0.5,1.0,2.0] for cases [LES\_B1, LES\_0, LES\_B2].  Quite evidently, the size of the ejections is halved for case LES\_B1 and doubled for case LES\_B2, compared to the baseline LES\_0.   Therefore we conjecture that the vortex rings are generated by the integral-scale structures in the original IHT field.


\begin{figure}
     \centering
     \begin{overpic}[width=0.75\textwidth]{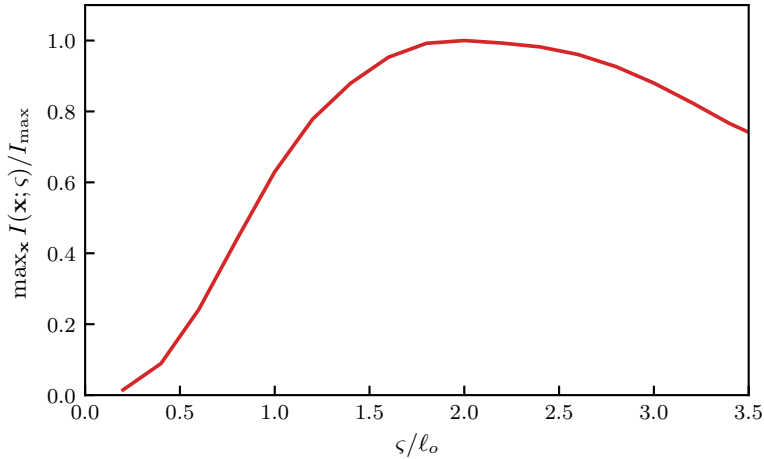}
     \end{overpic}
     \caption{ \small Maximum Gaussian weighted impulse over $\vx$, $\max_{\mathbf{x}} I(\mathbf{x};\varsigma)$ as a function of `width' $\varsigma$ for LES\_0 at $t=0$ (\protect \redline).}\label{fig:impulse}
 \end{figure}

We hypothesize that the ejections occur due to a local imbalance of impulse associated with the IHT field. Consider a Gaussian weighted impulse centered at point $\vx$, with a `width' $\varsigma$
\begin{align}
    \boldsymbol{I}(\vx;\varsigma) =  
        \int_{\R^3} e^{-\frac{|\vx-\vx'|^2}{2\varsigma^2}} \vu(\vx')\intdiff \vx'.
\end{align}
Figure \ref{fig:impulse} shows the maximum impulse $\boldsymbol{I}(\vx;\varsigma)$ over $\vx$, as a function of the width $\varsigma/ \ell_o$ at $t/t_\ell=0$, where $\ell_o$ is the initial integral scale. The maximum Gaussian weighted impulse reaches its maximum when $\varsigma / \ell_o$ is around $1.8$.

For points deep within the cloud, imbalance of the locally filtered impulse would simply result in complicated local vortex dynamics.  However, near the edge of the cloud, this imbalance, when pointed outwards, would eject vorticity out of the cloud.  In some sense, this process is universal as the scale is a property of the IHT field itself, and the net imbalance would create ejections near the edge of any region of IHT. This result also agrees with studies of TNTIs. It was discussed in \citet{townsend1980structure} that, while a wide range of turbulence scales affect the evolution of the turbulence boundary, the largest distortion at the TNTI is from the largest eddies in the turbulence.

\section{Concluding remarks}
\label{sec:conclusions}

We used DNS and LES to study a novel turbulent flow representing an isolated spherical region of turbulence evolving in free space. This flow is created by tiling a periodic IHT field in space and windowing it to be zero outside a spherical region.  The DNS is used to validate the LES, which is in turn used to study the long-time evolution of the turbulence.

The flow exhibits aspects of both homogeneous turbulence, deep within the sphere, as well as inhomoengoues turbulence near the TNTI.
For strictly homogeneous turbulence, a spectrum of either the Saffman $k^2$ type or the Batchelor $k^4$ type determines the kinetic energy decay rate and the the integral scale growth rate. For spherical region of turbulence we showed that both types of initial conditions can also be created.  For the cloud, we confirm, by comparing spectra on spherical shells of different radii from the initial center, that the turbulence remains locally homogeneous deep within the cloud.  However, the resulting long-term decay of the kinetic energy and the growth of the integral scale are similar in both cases, and closer to the predictions of the Saffman theory. This may be related to an observed growth in the Loitsyansky integral, but which is assumed constant in the Batchelor characterization of the turbulence.  At least through about 400 eddy turnover times, there is little difference in the shape of the respective spectra between the two cases for $k$ values near the inertial scale, and it appears that the integral scale is relatively unaffected by the behavior at very low $k$, whether $k^2$ or $k^4$.  In any event, the spectrum at these wavenumbers  is controlled by the initial conditions and may not be universal.  Finally,  we defined a mean radius of the turbulence cloud in terms of its velocity moments, and showed the turbulence gives rise to a similar growth of radius as of the integral scale.

The spherical region of turbulence is bounded by a TNTI that evolves into distinct large-scale features.  By varying each of the three independent nondimensional parameters controlling the cloud, we find that the structures are related to the (initial) integral scale of the IHT field.  The TNTI features include vortex rings that are ejected from the cloud.  We hypothesize that this evolution is associated with an imbalance in specific impulse over the integral scale, which, near the TNTI, gives rise to the vortex rings.

\section*{Acknowledgements}
This work was supported by the ONR grant No. N00014-16-1-2734 and the AFOSR/UCLA grant No. FA9550-18-1-0440.

\clearpage

\appendix
\section{Energy spectrum of inhomogeneous turbulent flows}\label{sec:appendix_total_spectrum}
Ambiguities arise in interpreting the (spatial) energy spectrum of inhomogeneous turbulent flows.  The term `spectrum' itself can be used in two different ways.  For a function, the spectrum can refer to the magnitude of its Fourier transform and gives information about the scales present in the function. For a random process, on the other hand, the spectrum represents a statistical statement about how energy is distributed amongst scales {\it on average}.  Turbulence is generally thought to be random in the sense (e.g. \cite{Pope_2001}) that any realization (e.g. specific initial condition) cannot be predicted with certainty from any other; only through an average of a sufficient number of realizations of the random process can we make statements about the likely properties of any.  In what follows, we interpret the term spectrum in this latter sense.  We discuss approximations we make in order to estimate the spectrum for the spherical region of turbulence under the approximation that the turbulence is locally homogeneous deep within the sphere.

\subsection{General definitions}

The {\it spatial} energy spectrum for an inhomogeneous flow can be formulated from the two-point velocity covariance tensor (written at some moment in time, and we suppress the temporal dependence in what follows)
\begin{equation}
 R_{ij}\left( {\bf x}, {\bf r} \right) = \mathbb{E} \left[ u_i({\bf x}) u_j({\bf x} + {\bf r} )\right],
 \label{eq:covar}
\end{equation}
where $\mathbb{E}$ is the expectation (ensemble average).  Note that the result depends on both the position in the flow ${\bf x}$ and the separation vector ${\bf r}$ between the observations. The Fourier transform (in the generalized sense) of $R_{ij}$ over the separation vector gives the cross-spectral density tensor
\begin{equation}
    S_{ij}({\bf x}, {\bf k}) = \frac{1}{8 \pi^3} \int_{\mathbb{R}^3} R_{ij} \left( {\bf x}, {\bf r} \right) e^{-i {\bf k} \cdot {\bf r}} \intdiff {\bf r}.
\end{equation}
where the integral is over the separation vector.

The resulting local kinetic energy spectrum (per unit volume) is
\begin{equation}
 \widehat{E}({\bf x}, {\bf k}) = \frac{1}{2} S_{ii}({\bf x}, {\bf k}).
\end{equation}
We can define a total kinetic energy spectrum of a domain $\Omega$ by integration,
\begin{equation}
 \widetilde{E}({\bf k}) = \int_\Omega \widehat{E}({\bf x}, {\bf k}) \intdiff {\bf x}, \label{eq:totale}
\end{equation}
and, by Parseval, the kinetic energy in the entire flow is
$\int \widetilde{E}({\bf k}) \intdiff {\bf k}$.
For turbulence that is homogeneous in one or more directions, discussed in more detail below, the integral diverges and the total energy is not defined.  However, in that case the local kinetic energy is also uniform in the homogeneous directions, and it is sufficient to speak of the energy per unit volume.

For future reference, using linearity of the expectation operator, we may also write
\begin{equation}
 \widetilde{E}({\bf k}) =\frac{1}{16 \pi^3}   \mathbb{E} \left[ \int_{\mathbb{R}^3}   \int_{\mathbb{R}^3} u_i({\bf x}) u_i({\bf x} + {\bf r} ) e^{-i {\bf k} \cdot {\bf r}} \intdiff {\bf x}\intdiff {\bf r} \right]. \label{eq:almost_greg}
\end{equation}
This form of the spectrum is often written without the expectation operator, but it then refers to the spectrum of a deterministic velocity field rather than that of an underlying random process. Evaluating it with the expectation requires an ensemble of realizations for the general case of inhomogeneous turbulence.

\subsection{Homogeneous and locally homogeneous turbulence}

Depending on additional hypotheses on the structure of the turbulence, different averaging procedures can be employed to determine a spatial or temporal spectrum in place of the ensemble average over realizations.  For example, if the turbulence is hypothesized as ergodic-stationary, then the ensemble average can be replaced by a sufficiently long time average over a single realization.  Likewise, if the turbulence is hypothesized as ergodic-homogeneous, then a spatial average over any or all homogeneous directions can be used.  For example, for the fully (all 3 directions) homogeneous case, we may write
\begin{align}
    R_{ij} = R_{ij}({\bf r}) =  \lim_{V \rightarrow \mathbb{R}^3} \frac{1}{V} \left( \int_{V} u_i({\bf x}) u_j({\bf x} + {\bf r} ) \intdiff {\bf x} \right). \label{homog}
\end{align}
The resulting cross-spectral density and local energy spectrum will likewise only be functions of the separation or wavenumber vectors, respectively, i.e. $S_{ij} = S_{ij}({\vr})$, $\widetilde{E} = \widetilde{E}({\vk})$.  Recall that total energy spectrum $\widetilde{E}(\vk)$ is infinite (undefined) in this case, since the integral over all space diverges.

Provided that the turbulence is {\it locally} homogeneous (or homogeneous plus isotropic) over a lengthscale $L$ such that $L \gg l$ ($l$ the integral scale) then the volume averaging can be performed locally.  Define a region $\Omega$ centered about ${\bf x}$ with scale $\Omega \sim L^3$, and define
\begin{align}
    \bar{R}_{ij}({\bf x}, {\bf r}) =  \frac{1}{ | {\Omega(x)} | } \left( \int_{\Omega(x)} u_i({\bf x}') u_j({\bf x}' + {\bf r} ) d {\bf x}' \right). \label{eq:local_homog}
\end{align}
We expect $\bar{R}_{ij}$ to be a constant over the region of homogeneity (except close to its edge).

\subsection{Estimating $\widetilde{E}(\vk)$ for the spherical cloud of turbulence}

We hypothesize that the turbulence is locally homogeneous over a region deep within the sphere of turbulence.  To apply this concept to the spherical cloud, we begin by breaking up the volume in equation \eqref{eq:almost_greg} into 3 parts: an inner region ($\Omega_{<R}$) deep in the sphere where we will assume local homogeneity, a transition region ($\Omega_{\sim R}$) near the turbulent/irrotational interface, and an outer, irrotational region ($\Omega_{>R}$)
\begin{align}
 \widetilde{E}{(\bf k}) & =\frac{1}{16 \pi^3}  \int_{\mathbb{R}^3}  \left[  \int_{\Omega_{<R}} + \int_{\Omega_{\sim R}}  + \int_{\Omega_{>R}} \right] \mathbb{E} \left[ u_i({\bf x}) u_i({\bf x} + {\bf r} )\right]    \, e^{-i {\bf k} \cdot {\bf r}} \intdiff  {\bf x}  \intdiff {\bf r}.
\end{align}
With local homogeneity over $\Omega_{<R}$, we insert equation~\eqref{eq:local_homog}, which is constant with ${\bf x}$ over this region, into the first integral, and obtain
\begin{align}
 \widetilde{E}({\bf k}) & =\frac{1}{16 \pi^3}  \int_{\mathbb{R}^3}   \int_{\Omega_{<R}}u_i({\bf x}) u_i({\bf x} + {\bf r} )  \, e^{-i {\bf k}  \cdot {\bf r}} \intdiff {\bf x} \intdiff {\bf r} \nonumber \\
 & + \frac{1}{16 \pi^3} \left[ \int_{\Omega_{\sim R}}  + \int_{\Omega_{>R}} \right] \mathbb{E} \left[ u_i({\bf x}) u_i({\bf x} + {\bf r} )\right]  \, e^{-i {\bf k}  \cdot {\bf r}}  \intdiff {\bf x} \intdiff {\bf r} \nonumber \\
 & =\frac{1}{16 \pi^3}  \int_{\mathbb{R}^3}  \int_{\mathbb{R}^3}  u_i({\bf x}) u_i({\bf x} + {\bf r} )    \, e^{-i {\bf k}  \intdiff {\bf x}  \cdot {\bf r}}   \intdiff {\bf r} + E^{\prime}_{\Omega_{\sim R}} + E^{\prime}_{\Omega_{>R}} ({\bf k}), \label{eq:greg_witherror}
\end{align}
where the remainder terms are of the form of a {\em difference} between the ensemble average and one realization, i.e.
\begin{align}
 \widetilde{E}^{\prime}({\vk})_{\Omega}  & = \frac{1}{16 \pi^3}   \int_{\mathbb{R}^3}  \int_{\Omega} \left( \mathbb{E} \left[ u_i({\vx}) u_i({\vx} + {\vr} ) \right] - u_i({\vx})  u_i({\vx} + {\vr} )   \right)   \, e^{-i {\vk} \cdot {\vr}} \intdiff {\vx}  \intdiff{\vr}.
\end{align}
In breaking up the integral in this way, we highlight that we can integrate the velocity field from a single simulation over free space and obtain the correct ensemble-averaged spectrum up to a statistical error associated only with the difference between one realization and the ensemble average only over the transition and outer regions.

Regarding the transition region, the contribution to the overall energy scales with the volume of this region, $4 \pi \sigma R^2$, where $\sigma$ is the width of the transition region.  By making the initial sphere large compared to $\sigma$ and the correlation length (integral scale), $\ell(x)$, this error can, at least in principle, be made indefinitely small compared to the first term.

Regarding the outer region, the irrotational velocity field decays at least as fast as $|{\bf x}|^{-3}$ when the initial impulse is nonzero.  For wavenumber $k$ not too small, we expect this to only produce a small contribution to the total energy spectrum.  However, as $k \rightarrow 0$, this term will eventually {\it dominate} the spectrum, and the behavior at low $k$ will be $O(k^2)$ provided the initial impulse is nonzero.  Indeed, we therefore do not expect the low wavenumber spectrum to be universal as it depends on how much impulse there is in the initial condition, which, as described in \S~\ref{sec:initial_conditions}, is arbitrary and can be contrived, with little effect on the resulting turbulence, to have any value (including zero).  When the impulse is zero, the resulting $k^4$ spectrum may be universal over a broader range of low wavenumbers, but is still contrived as $k \rightarrow 0$.

This discussion has strong, but unsurprising implications about whether the low wavenumber spectrum of {\it any} turbulent flow can be considered to be universal.  An alternative interpretation is that it is associated with the initial/boundary conditions and can be arbitrarily manipulated independently of the turbulence behavior at smaller scales.  In any event, it is clear that, in the present simulations, the low wavenumber behavior is wholly controlled by the (arbitrary) initial condition.  At sufficiently long time, after the turbulence has substantially decayed, the error terms above will eventually dominate the spectrum, resulting in a (in the case of finite impulse), a fat vortex ``puff'' whose properties are solely related to the initial condition.

If the error terms are neglected, equation~\eqref{eq:greg_witherror} is identical to equation~\eqref{eq:almost_greg}, but without the expectation.  It has been used before to express the energy spectra of deterministic velocity fields but it's equivalence (to within the error) to the ensemble-averaged spectrum of a random process, under local homogeneity, has not to our knowledge been reported elsewhere. It is interesting that the error vanishes like the ratio of the volume of inhomogeneous turbulence to volume of homogeneous turbulence.

\section{Spectrum: analytical and computational  details}\label{sec:appendix_total_spectrum_B}
For $kR \gtrsim 1$, equation~(\ref{eq:greg}) is evaluated using the three-dimensional vorticity form
\begin{align}
 \widetilde{E}(\vk)
 & =\frac{1}{16 \pi^3} \int_{\R^3}  \int_{\R^3}  \vu(\vx)\cdot \vu(\vx')\,  e^{-i \vk \cdot (\vx'-\vx)} \intdiff \vx  \intdiff \vx' \label{eq:blowup_vel}\\
 & =\frac{1}{16 \pi^3} \int_{\R^3}  \int_{\R^3}  \frac{1}{|\vk|^2} \vomega(\vx)\cdot \vomega(\vx')\,  e^{-i \vk \cdot (\vx'-\vx)} \intdiff \vx  \intdiff \vx' \nonumber \\
&=
    \frac{1}{16 \pi^3 \mathbf{|k|}^2}  \mathcal{F}\{\vomega(\mathbf{x})\} \cdot  \overline{\mathcal{F}\{\vomega(\mathbf{x})\} }, \label{eq:blowup}\\
E(k) &= \int_{S_k} \widetilde{E}(\vk) \intdiff S_k,
\end{align}
where $\mathcal{F}\{ \cdot \}$ denotes the Fourier transform and $S_k$ denotes a spherical shell of radius $k$. Fast Fourier transform is used to efficiently evaluate (\ref{eq:blowup}) and zero-padding is applied to attenuate the effect from the spurious periodicity.

For $kR \lesssim 1$, formula (\ref{eq:blowup}) must be evaluated carefully to avoid numerical singularity.  For these values, we expand the integral in a Taylor series about $k=0$ to obtain
\begin{align}
E(k)
    &= \frac{1}{4\pi^2}\sum_{i=1}^{\infty} (-1)^i\frac{1}{(2i+1)!} k^{2i} \int_{\mathbb{R}^3} \mathbf{|r|}^{2i}
    \mathcal{F}^{-1}\left\{\mathcal{F}\left\{\vomega\right\}\cdot \overline{\mathcal{F}\left\{\vomega\right\}}\right\}(\mathbf{r}) \intdiff \mathbf{r}, \label{eq:vor_expansion}
\end{align}
which already uses the relation that $\int \vomega(\vx) \intdiff \vx=0$.
For all results presented here, up to 10 terms are used to yield accurate spectrum for the low wavenumber limit.

A more common form in terms of the velocity field for the low wavenumber limit can be derived by expanding equation~\eqref{eq:blowup_vel}
\begin{align}
E(k)
    &= \frac{1}{4\pi^2}\sum_{i=0}^{\infty} (-1)^i\frac{1}{(2i+1)!} k^{2i+2} \int_{\mathbb{R}^3} \mathbf{|r|}^{2i}
    \mathcal{F}^{-1}\left\{\mathcal{F}\left\{\vu\right\}\cdot \overline{\mathcal{F}\left\{\vu\right\}}\right\}(\mathbf{r}) \intdiff \mathbf{r}.
\end{align}
Comparing the corresponding terms in equation~\eqref{eq:blowup_vel} and \eqref{eq:blowup} gives another relation
\begin{align}
    \int_{\R^3}\int_{\R^3} r^{2p+2} \vomega(\vx)\cdot\vomega(\vx+\vr)\intdiff \vx \intdiff\vr =  -(2p+2) (2 p+3)  \int_{\R^3} \int_{\R^3} r^{2p} \, \vu(\vx)\cdot\vu(\vx+\vr) \intdiff \vx \intdiff \vr.
\end{align}

Lastly one can show the $k^2$ term in equation~\eqref{eq:vor_expansion} is related to the total vorticity impulse $\boldsymbol{J}_\omega$ through
\begin{align}
    \int_{\R^3}\int_{\R^3} r^{2} \vomega(\vx)\cdot\vomega(\vx+\vr)\intdiff \vx \intdiff\vr =4 |\boldsymbol{J}_\omega|^2,
\end{align}
where
\begin{align}
     \boldsymbol{J}_\omega=\frac{1}{2}\int_{\R^3} \vx \times \vomega(\vx) \intdiff \vx.
\end{align}

\newpage
\bibliography{main}
\bibliographystyle{jfm}

\end{document}